\documentclass[final,5p,times,twocolumn]{elsarticle}

\usepackage[utf8]{inputenc}
\usepackage{amsmath,amssymb}
\usepackage{xcolor}
\usepackage{hyperref}
\usepackage[capitalize]{cleveref}

\journal{Physics Letters B}

\biboptions{numbers,sort&compress} 
\begin{document}

\begin{frontmatter}

\title{
\vskip-2cm\hfill December 2025\\[1cm]
On the Charm Contribution to the Muon $g-2$ Light-by-Light}

\author[a]{Johan Bijnens}
\author[b]{Nils Hermansson-Truedsson}
\author[c]{Antonio Rodr\'{\i}guez-S\'anchez}

\address[a]{Division of Particle and Nuclear Physics, Department of Physics, Lund University,\\
Box 118, SE 221-00 Lund, Sweden}

\address[b]{Higgs Centre for Theoretical Physics, School of Physics and Astronomy,\\
The University of Edinburgh, James Clerk Maxwell Building, Peter Guthrie Tait Road,\\
Edinburgh EH9 3FD, United Kingdom}

\address[c]{Departamento de F\'{\i}sica, Universidad de Castilla-La Mancha,\\
Avenida de Carlos III, s/n, 45004 Toledo, Spain}

\begin{abstract}
We combine existing perturbative results to show that a precise analytic determination of the charm-quark contribution to the hadronic light-by-light (HLbL) part of the muon anomalous magnetic moment is possible. Working in the $\overline{\mathrm{MS}}$ scheme, we include the NLO $\mathcal{O}(\alpha_s)$ correction, which significantly reduces the residual renormalization-scale dependence and the perturbative uncertainty. 
Our final result is $a_{\mu}^{\mathrm{HLbLc}}=3.65(25)\times 10^{-11}$, in good agreement with recent lattice determinations.
\end{abstract}


\end{frontmatter}

%% main text

\section{Introduction}
One of the most studied observables in particle physics in recent years is the anomalous magnetic moment of the muon~\cite{Muong-2:2025xyk,Aoyama:2020ynm,Aliberti:2025beg}. Much of the recent activity has focused on the hadronic light-by-light (HLbL) contribution, using phenomenological approaches~\cite{Colangelo:2015ama,Masjuan:2017tvw,Colangelo:2017fiz,Hoferichter:2018kwz,Eichmann:2019tjk,Bijnens:2019ghy,Leutgeb:2019gbz,Cappiello:2019hwh,Masjuan:2020jsf,Bijnens:2020xnl,Bijnens:2021jqo,Danilkin:2021icn,Stamen:2022uqh,Leutgeb:2022lqw,Hoferichter:2023tgp,Hoferichter:2024fsj,Estrada:2024cfy,Ludtke:2024ase,Deineka:2024mzt,Eichmann:2024glq,Bijnens:2024jgh,Hoferichter:2024bae,Holz:2024diw,Cappiello:2025fyf} and lattice QCD calculations~\cite{Blum:2019ugy,Chao:2021tvp,Chao:2022xzg,Blum:2023vlm,Fodor:2024jyn}. Although this piece is subleading, being suppressed by an additional factor $\mathcal{O}\!\left[(m_\mu/(2m_c))^2\right]$, there has also been interest in the charm-quark contribution in both communities, making it a useful benchmark for comparing QCD techniques. For the $\tau$ lepton, the analogous contribution is less suppressed and has recently been discussed in Ref.~\cite{Hoferichter:2025fea}. 

The main focus of this short letter is to improve the perturbative evaluation of this charm contribution, presented in \cref{sec:charmlbl}, where we also discuss the electron and tau cases. Comparisons with previous results and our conclusions are given in \cref{sec:discussion}.

%Need to discuss \cite{Hoferichter:2025fea}

\section{Charm contribution to the light-by-light}
\label{sec:charmlbl}

The charm mass is known to be sufficiently large for perturbative QCD series to be well behaved in the first few orders. For Euclidean momenta, away from hadronic singularities, the charm mass acts as a regulator and allows for perturbative evaluations of Green functions even for $Q^{2}< m_{c}^2$. For example, see Figs.~7 and 8 of Ref.~\cite{Davier:2023hhn} for the two-point vector function. In this regime, the perturbative series take the form
\begin{equation}\label{eq:PiHeavy}
\Pi(Q)=\sum_j \overline{C}_{j}(\mu)\, z^j(\mu) \, ,
\end{equation}
where 
\begin{equation}\label{eq:CjCoefficients}
\overline{C}_{j}(\mu)=\sum_{n}\overline{C}_{j}^{(n)}(\mu)\left(\frac{\alpha_s(\mu^2)}{\pi}\right)^n  \, ,
\quad
z(\mu)=-\frac{Q^2}{4 m_c^2(\mu^2)} \, ,
\end{equation}
and power corrections are known to be suppressed in this regime~\cite{Novikov:1977dq}. 

For the evaluation of the charm contribution to the light-by-light part of the muon $g-2$, the effective low-energy scale for the expansion is $m_\mu$. The lowest-order term was obtained long ago from the muon-loop contribution to the electron $g-2$. This calculation was performed fully analytically as an expansion in $m_\mu/m_c$ in Ref.~\cite{Kuhn:2003pu}, with the first two terms already given in Ref.~\cite{Laporta:1992pa}. These results were used in Refs.~\cite{Aoyama:2020ynm,Aliberti:2025beg,Colangelo:2019uex} to estimate the charm contribution, assigning an uncertainty based on the bound $c\bar{c}$ states. The equal mass case is known fully analytically \cite{Laporta:1991zw}.

\begin{figure}[tb]
    \centering
    \includegraphics[width=0.98\linewidth]{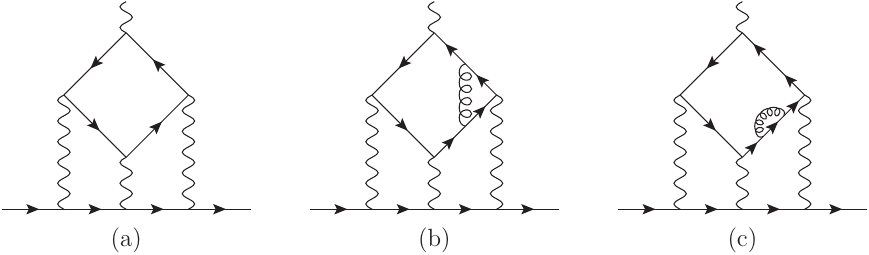}
    \caption{Types of diagrams that contribute. (a) Leading-order diagram, (b) crossed-gluon diagram, (c) diagram with mass renormalization.}
    \label{fig:diagrams}
\end{figure}

A representative lowest-order diagram is shown in \cref{fig:diagrams}(a). The evaluation to first order in $m_\mu/m_c$ is \cite{Laporta:1992pa,Boughezal:2011vw}
\begin{align}
\label {eq:LO}
A_\mu^{\text{HLbLc LO}} &= \left(\dfrac{\alpha}{\pi}\right)^3\left(\dfrac{3}{2}\, \zeta(3)-\dfrac{19}{16}\right)
N_c q_c^4\, \dfrac{m_\mu^2}{M_c^2}\,.
\end{align}
where $q_c = 2/3$ is the charm-quark electric charge.
The first correction in $\alpha_s$ comes from the type of diagrams shown in \cref{fig:diagrams}(b,c). These diagrams also enter the $\tau$-loop contribution to the muon $g-2$ and the muon-loop contribution to the electron $g-2$. The analytic expression to first order in $m_\mu/m_c$ was obtained in Ref.~\cite{Boughezal:2011vw} in a different context and is given by
\begin{align}
\label{eq:NLO}
 A_\mu^{\text{HLbLc NLO}}  &= A_\mu^{\text{HLbLc LO}} 
\dfrac{\alpha_s}{\pi} C_F\dfrac{\Delta_1}{\Delta_0} \, ,
 \nonumber\\
 \Delta_0 &= \dfrac{3}{2}\, \zeta(3)-\dfrac{19}{16} \, ,
 \nonumber\\
\Delta_1 &= 
-\frac{473  \pi^2}{1080}\ln^22
+\frac{52 \pi^2}{405}\ln^32
-\frac{42853 \pi^4}{259200}
\nonumber\\
&\quad+\frac{5771\pi^4}{32400}\ln2
+\frac{473}{1080}\ln^42
-\frac{52}{675}\ln^52
\nonumber\\
&\quad-\frac{8477}{2700}  
+\frac{473}{45}a_4
+\frac{416}{45}a_5
+\frac{34727 \zeta_3}{2400}
\nonumber\\
&\quad
-\frac{23567 \zeta_5}{1440}.
\end{align}
The symbols used in \cref{eq:NLO} are $C_F = 4/3$, $a_{4,5} = {\rm Li}_{4,5}(1/2)$ the polylogarithmic constants, and $\zeta_{3,5}$ the Riemann zeta values at 3 and 5.
Numerically, $C_F \Delta_1/\Delta_0$ evaluates to 
$3.85174$. 

The result \cref{eq:NLO} of Ref.~\cite{Boughezal:2011vw} was obtained in the on-shell scheme for constituent-quark masses. We denote this mass by $M_c$. 
However, it is well known that the pole mass suffers from a renormalon ambiguity~\cite{Beneke:1994sw}, leading to poorly behaved perturbative series. This can be avoided by writing the expansions in terms of the $\overline{\mathrm{MS}}$ masses~\cite{ParticleDataGroup:2024cfk}. Indeed,
it is known that typically expansions in the $\overline{\mathrm{MS}}$ scheme are known to converge better~\cite{Maier:2007yn}. To do this, we convert the pole mass to the $\overline{\mathrm{MS}}$ mass. The relation needed at this order was calculated long ago~\cite{Tarrach:1980up,Gray:1990yh}
\begin{align}
\label{eq:mass}
    M_c = m_c(M_c)\left(1+\dfrac{4}{3}\dfrac{\alpha_s}{\pi}\right) \, .
\end{align}
In addition one needs to take into account the running of the quark mass.
Taking all effects into account leads to 
\begin{align}
\label{eq:LO2}   
A_\mu^{\text{HLbLc LO, }\overline{\textrm{MS}}} &= \left(\dfrac{\alpha}{\pi}\right)^3\left(\dfrac{3}{2}\, \zeta(3)-\dfrac{19}{16}\right)
N_c q_c^4\, \dfrac{m_\mu^2}{m_c(\mu)^2}\,,
\nonumber\\
 A_\mu^{\text{HLbLc NLO, }\overline{\textrm{MS}}}  &= A_\mu^{\text{HLbLc LO, }\overline{\textrm{MS}}}  
\dfrac{\alpha_s}{\pi}
\nonumber\\ &~~~~~\times
\left( C_F\dfrac{\Delta_1}{\Delta_0}-\dfrac{8}{3}
+ C \log\left(\dfrac{m_c(\mu)}{\mu}\right)\right) \, .
\end{align}
The coefficient $C=4$ is determined from the requirement that, to order $\alpha_s$, the result is $\mu$-independent.

\begin{figure}[tb]
    \centering
    \includegraphics[width=1\linewidth]{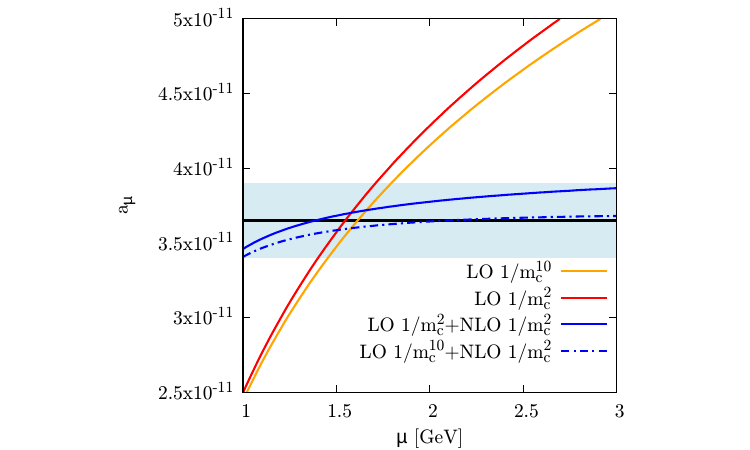}
    \caption{The contribution from the charm quark to $a_\mu^{\textrm{HLbL}}$. LO $1/m_c^{10}$ is the result at lowest order including higher orders in $m_\mu/m_c$ \cite{Kuhn:2003pu}. LO $1/m_c^2$ is the result of only the lowest order in $m_\mu/m_c$. LO $1/m_c^{2}$+NLO $1/m_c^{2}$ is our full result with the central values of \cref{eq:inputs} as input, i.e.~leading in $m_\mu/m_c$ but containing $\alpha_s$ corrections. LO $1/m_c^{10}$+NLO$1/m_c^{2}$ has the higher orders in $\mu_\mu/m_c$ included in the LO part.}
    \label{fig:results}
\end{figure}

For numerical input we use the values
\begin{align}
\label{eq:inputs}
    m_c(m_c) &= 1.2730\pm0.0046~\text{GeV}\,,
    \nonumber\\
    \alpha_s(m_c) &= 0.3876\pm0.0118\,.
\end{align}
These values are taken from Ref.~\cite{ParticleDataGroup:2024cfk} and evolved down to $m_c$ using five-loop running, with a matching scale between $m_b$ and $2m_b$. 
The results are plotted in \cref{fig:results}. We run $\alpha_s$ and $m_c$ from $m_c$ to $\mu$ at five-loop order with four active quark flavours, using both CRunDec \cite{Herren:2017osy} and our own implementation, finding agreement. As can be seen, the LO  $1/m_c^{2}$+NLO $1/m_c^{2}$ result as a function of the subtraction scale $\mu$ is much better behaved.
Using instead LO $1/m_c^{10}$+NLO $1/m_c^{2}$ does not change the conclusion.
\begin{figure}[tb]
    \centering
    \includegraphics[width=1\linewidth]{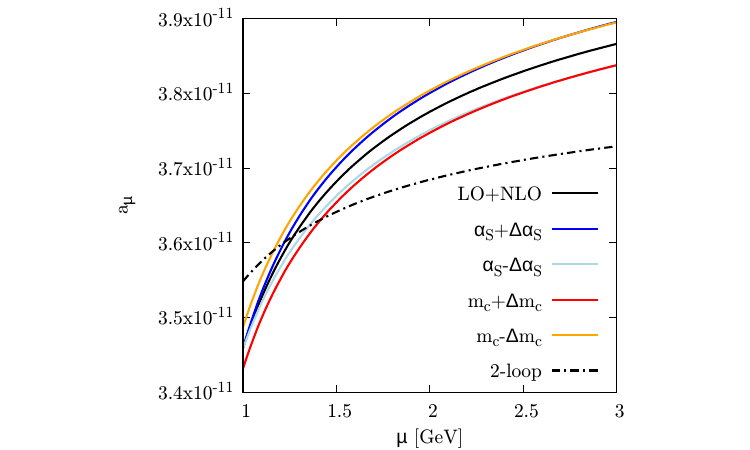}
    \caption{The contribution from the charm quark to $a_\mu^{\textrm{HLbL}}$. The result is LO+NLO while varying the input value of the charm quark mass and $\alpha_s$ within the errors given in \cref{eq:inputs}. The last curve is done with the same input but 2-loop running of $m_c$ and $\alpha_s$.} 
    \label{fig:results2}
\end{figure}
The variation of the result with $\alpha_s$ and $m_c$ within the bounds of \cref{eq:inputs} is shown in \cref{fig:results2}. This variation is much smaller than the changes from varying the subtraction scale $\mu$. A reasonable conclusion from the graphs shown is that the charm contribution is given by
\begin{align}
\label{eq:final}
  a_\mu^{\text{HLbLc}} = 3.65(25) \times 10^{-11}\,.  
\end{align}
This corresponds to the black line with shaded background in \cref{fig:results} and covers the entire area shown in \cref{fig:results2}.
This represents a significantly improved error compared to the result used in \cite{Aoyama:2020ynm,Aliberti:2025beg,Colangelo:2019uex} of $3(1) \times 10^{-11}$ in the analytical and dispersive section. 

Knowledge of higher orders in these perturbative series would help to further reduce the associated uncertainties. While these terms are not known, we can support our uncertainty estimates by applying the same criterion to other RGE-invariant series for which higher-order coefficients are available~\cite{Maier:2007yn,Maier:2009fz}. The first moment of the non-singlet vector two-point function is known up to four loops and, including subsequent perturbative orders, one finds,
\begin{equation}
\label{eq:PiV}
\Pi_{V}(Q\ll m_c)\propto (1+0.295+0.036-0.010)\,z(m_c) \, ,
\end{equation}
while the same error method and relative error gives $\Pi_{V}\propto (1.295\pm 0.089)\, z(m_c)$, thus effectively covering those higher-order terms. A similar behavior can be obtained for the axial-vector case and the higher moments.

The result in \cref{eq:final} can also be compared with what one would obtain using an on-shell mass. Then one uses \cref{eq:LO} and \cref{eq:NLO}. For a fixed on-shell mass $M_c=1.49$~GeV and $\alpha_s(M_c)=0.3519$,
we obtain 
\begin{align}
  a_\mu^{\textrm{HLbLc pole}} &= 
  2.30\times 10^{-11}~(\mathrm{LO}~1/M_c^2)\,,\nonumber\\
  &=2.25\times 10^{-11} ~(\mathrm{LO}~1/M_c^{10})\,,\nonumber\\
  &=3.29\times 10^{-11}~(\mathrm{LO}~ 1/M_c^2+\mathrm{NLO}~ 1/M_c^2)\,. 
\end{align}
The labels have the same meaning as in \cref{fig:results}.
Here one notices that we obtain a very large $\alpha_s$ correction, much larger than in \cref{fig:results}. Assuming that higher-order corrections are also large,\footnote{For comparison, the equivalent of \cref{eq:PiV} in the on-shell scheme, taking $\alpha_s(M_c)=0.3519$, is $\Pi_{V}(Q\ll M_c)\propto (1+0.567+0.391+0.483)\,z(M_c)$, thus having very large higher order corrections.}
this is in reasonable agreement with \cref{eq:final}.

\begin{figure}[tb]
    \centering
    \includegraphics[width=1\linewidth]{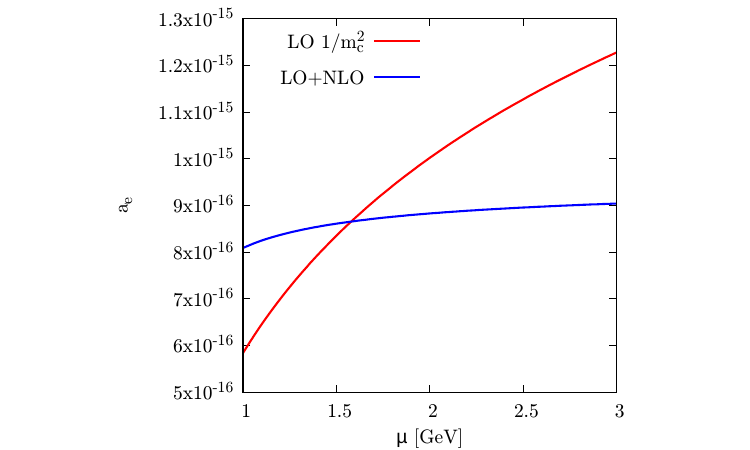}
    \caption{The charm quark contribution to the electron HLbL.}
    \label{fig:resultselectron}
\end{figure}

\begin{figure}[tb]
    \centering
    \includegraphics[width=1\linewidth]{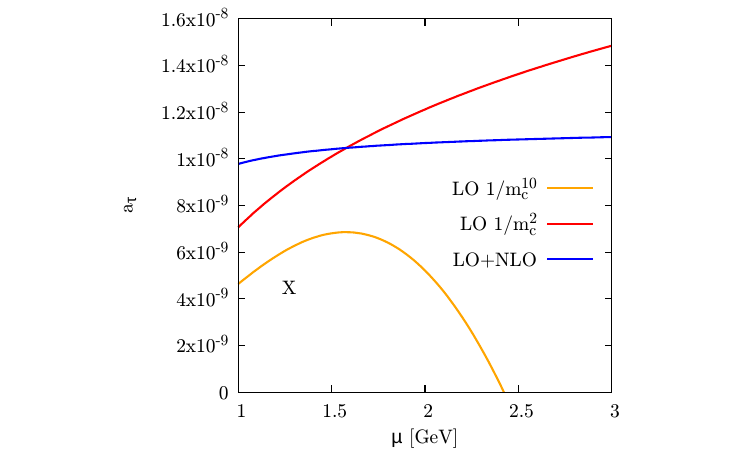}
    \caption{The charm loop results for the tau HLbL. The X indicates the full LO result at $m_c=1.273$.}
    \label{fig:tau}
\end{figure}

The same method can be used for the electron. In this case, the difference between the lowest-order result in the quark-mass expansion and the higher-order result is indistinguishable. The result is shown in \cref{fig:resultselectron}.
This leads to
\begin{align}
    a_e^{\textrm{HLbLc}} = 8.5(5)\times 10^{-16}\,.
\end{align}
The tau can be treated in the same way. This has been evaluated in Ref.~\cite{Hoferichter:2025fea} for the full quark-loop at $m_c(m_c)$ with
$a_\tau^{\textrm{HLbLc}} = 4.5\times 10^{-9}$. Our results are at the leading order in $1/m_c^2$ and are shown in \cref{fig:tau}.
The stabilization as a function of the subtraction scale $\mu$ is similar to the previous cases, but keeping only the leading term in $1/m_c^2$ is not necessarily a good approximation. We therefore do not quote a final number for this case.

\section{Final discussion}
\label{sec:discussion}
We have incorporated the $\alpha_s$ correction to the charm-loop contribution to the muon HLbL $g-2$ in the $\overline{\mathrm{MS}}$ scheme, translating the $\alpha_s$ corrections to (model-dependent) constituent quark contributions computed in Ref.~\cite{Boughezal:2011vw}. We have argued that this leads to better-controlled perturbative uncertainties. Our final number is 
\begin{equation}
a_{\mu}^{\mathrm{HLbLc}}=3.65(25)\times 10^{-11} \, ,
\end{equation}
which can be compared to $a_{\mu}^{\mathrm{HLbLc}}=3(1)\times 10^{-11}$ from Ref.~\cite{Aliberti:2025beg}, obtained without incorporating gluon corrections, or to the slightly different value of Ref.~\cite{Hoferichter:2025fea}, $a_{\mu}^{\mathrm{HLbLc}}=2.7(8)\times 10^{-11}$, which attempts to consider corrections to the quark loop with a different method.

The charm quark contribution has also been calculated using lattice QCD. There are two full calculations. The Mainz/CLS collaboration finds \cite{Chao:2022xzg}
\begin{align}
a_\mu^{\textrm{HLbLc}} = 2.8(5) \times 10^{-11} \,.
\end{align}
The final result of the BMW collaboration is \cite{Fodor:2024jyn}
\begin{align}
    a_\mu^{\textrm{HLbLc}} = 3.73(26) \times 10^{-11} \,.
\end{align}
The BMWc result is in very good agreement with ours, while there is mild tension with the Mainz/CLS result.
These calculations also include estimates of the disconnected contribution. This does not appear in our result, since it starts contributing only at order $\alpha_s^2$, beyond the order we have considered, and also contains a nonperturbative part. This contribution was calculated by BMW and found to be
$-0.185(52)(04)$ consistent with the Mainz/CLS calculation. As this is a sizable shift, we can compare to the connected lattice QCD charm-quark result in isolation. The connected part of the BMWc the result is~\cite{Fodor:2024jyn}
\begin{align}
      a_\mu^{\textrm{HLbLc, conn.}} = 3.92(26) \times 10^{-11} \,.
\end{align}
The Mainz/CLS collaboration has~\cite{Chao:2022xzg}
\begin{align}
      a_\mu^{\textrm{HLbLc, conn.}} = 3.1(4) \times 10^{-11} \,.
\end{align}
As can be seen, our value is consistent within errors.

\section*{Acknowledgements}

This work has been supported by 
the Spanish Government (Agencia Estatal de Investigaci\'on MCIN/AEI/10.13039/501100011033) Grant No. PID2023-146220NB-I00. N.~H.-T.~is supported by the UK Research and Innovation, Science and Technology Facilities Council, grant number UKRI2426.

\bibliographystyle{elsarticle-num}  
\bibliography{refs}
\end{document}